\documentstyle[aps,epsfig,floats]{revtex}

\def\lsi{\raise0.3ex\hbox{$<$\kern-0.75em\raise-1.1ex\hbox{$\sim$}}}
\def\gsi{\raise0.3ex\hbox{$>$\kern-0.75em\raise-1.1ex\hbox{$\sim$}}}

\begin{document}
\draft
\twocolumn[\hsize\textwidth\columnwidth\hsize\csname
@twocolumnfalse\endcsname
\title{Possible 
detection of relic neutrinos and determination of 
their mass:\\ quantitative analysis$^\star$}
\author{
Z.~Fodor$^1$, S.D.~Katz$^1$ and A. Ringwald$^2$}
\address{$^1$Institute for Theoretical Physics, E\"otv\"os University, 
P\'azm\'any
1, H-1117 Budapest, Hungary\\
$^2$Deutsches Elektronen-Synchrotron DESY, Notkestr. 85, D-22607,
Hamburg, Germany}

\date{\today}
\maketitle

\vspace*{-4.0cm}
\noindent
\hfill \mbox{ITP-Budapest 569\ \ \ \ DESY 01-070}
\vspace*{3.8cm}

\begin{abstract} \noindent
We consider the possibility that a large fraction of the 
ultrahigh energy cosmic rays (UHECRs)  are decay products of Z bosons 
which were produced in the scattering of ultrahigh energy cosmic neutrinos
 (UHEC$\nu$s) on cosmological relic neutrinos (R$\nu$s).   We 
compare the observed UHECR spectrum with the one predicted in the 
above Z-burst scenario and determine the  mass of the heaviest
 R$\nu$ as well as the necessary UHEC$\nu$ flux via a maximum 
likelihood analysis. 
\end{abstract}

\vskip1.3pc]

\section{Introduction}

Big-bang cosmology predicts the existence of the cosmic microwave background 
radiation (CMBR) and a similar background of R$\nu$s 
with an average number density of 
$\langle n_{\nu_i}\rangle\approx 56$ cm$^{-3}$ per light 
neutrino species $i$ ($m_{\nu_i}<1$ MeV). However, the R$\nu$s have not been 
detected until now. 

A possibility for their detection was discussed some time 
ago:
the UHEC$\nu$ spectrum should have absorption dips at energies 
$\approx E_{\nu_i}^{\rm res} = M_Z^2/(2\,m_{\nu_i}) = 4.2\cdot 10^{21}$ eV  
(1 eV/$m_{\nu_i}$) due to resonant annihilation with the R$\nu$s  
into Z bosons of mass $M_Z$~\cite{W82,R93,Y97}.
Recently it was realized that the same annihilation mechanism might
already be visible in the UHECR spectrum~\cite{FMS99,W99} at energies
above the predicted Greisen-Zatsepin-Kuzmin (GZK) 
cutoff~\cite{G66,ZK66,BS00} 
ar\-ound $4\cdot 10^{19}$~eV. It was argued that the UHECRs above 
the GZK cutoff are mainly protons from Z decay.
 
This hypothesis was discussed in several 
papers~\cite{W98,Y99,GK99,BVZ00,PW01,FGDTK01}.
We report here on our recent 
quantitative investigation of the Z-burst scenario, where we
have determined the mass of the heaviest
R$\nu$ as well as the 
necessary UHEC$\nu$ flux via a maximum 
likelihood analysis~\cite{FKR01a}. 

Our comparison of the Z-burst scenario with the observed UHE\-CR spectrum
was done in four steps. First, we determined the probability
of Z production as a function of the distance from Earth.
Secondly, we exploited collider experiments to derive 
the energy distribution of the produced protons in the lab system.
Thirdly, we considered the propagation of the protons, i.\,e. we determined 
their energy losses due to pion and $e^+e^-$ production 
through scattering on the CMBR and due to their redshift. The last
step was the comparison of the predicted and observed spectra and 
the extraction of the mass of the R$\nu$ and the necessary UHEC$\nu$ flux.   

\section{Z-burst spectrum}

Our prediction of the contribution of protons 
from Z-
$\underline{\hspace{3cm}}$\\
$^\star${\footnotesize Talk to be presented 
at the 27th International Cosmic Ray Conference, Hamburg, Germany, 
August 7-15, 2001.}

\newpage\noindent 
bursts to the UHECR spectrum, for degenerate $\nu$ masses
($m_\nu\approx m_{\nu_i}$), can be summarized as
\begin{eqnarray}
\label{nu-flux}
\lefteqn{ j(E,m_\nu)=I\cdot F_Z^{-1} \cdot \sum_{i}
\int\limits_0^\infty {\rm d}E_p \int\limits_0^{R_0} {\rm d}r 
\int\limits_0^\infty {\rm d} \epsilon } 
\\[1ex] \nonumber &&
 F_{\nu_i}(E_{\nu_i},r)n_{\nu_i}(r)\sigma( \epsilon)
{\cal Q}(E_p)
\left(-\partial P(r,E_p,E) / \partial E\right),
\end{eqnarray}
where the total time and angle integrated detector area 
$I$ and the normalization factor 
$F_Z$, which is proportional to the sum of the $\nu$ fluxes at 
centre-of-mass (CM) energy 
$M_Z$, are determined later by the comparison with the UHECR data.   
$E$ is the energy of the protons arriving at Earth. 
Further important ingredients in our prediction~(\ref{nu-flux}) are: 
the UHEC$\nu$ fluxes $F_{\nu_i}(E_{\nu_i},r)$ at the resonant energy 
$E_{\nu_i}\approx E_{\nu}^{\rm res}$ and at distance $r$ to Earth, 
the number density 
$n_{\nu_i}(r)$ of the R$\nu$s, 
the Z production cross section $\sigma(\epsilon)$ at CM energy 
$\epsilon = \sqrt{2\,m_\nu\,E_{\nu_i}}$, 
the energy distribution ${\cal Q}(E_p)$ of the produced protons with energy 
$E_p$, 
and the probability $P(r,E_p,E)$ 
that a proton created at a distance $r$ with energy $E_p$ arrives
at Earth above the threshold energy $E$. 

The last three building blocks, $\sigma$, ${\cal Q}$, and $P$, are very well 
determined.  
At LEP and SLC millions of Z bosons 
were produced and their decays analyzed with extreme high accuracy. We used 
existing published~\cite{Akers94,Abreu95,Buskulic95,Abe99} and some improved 
unpublished~\cite{Abe01} data  to determine the 
proton momentum distribution ${\cal Q}(E_p)$.
Due to the large statistics, the uncertainties of our analysis
related to Z decay turned out to be negligible. 
Similarly, the CMBR is known to a high accuracy.  It plays the key role in the 
determination of the probability $P(r,E_p,E)$~\cite{BW99,FK01a}, which takes 
into 
account the fact that protons of extragalactic (EG) origin and energies above 
$\approx 4\cdot 10^{19}$ eV 
lose a large fraction of their energies~\cite{G66,ZK66}
due to pion and $e^+e^-$ production 
through scattering on the CMBR and due to their redshift.
$P(r,E_p,E)$, in the form as it has been calculated for a wide range
of parameters by two of the present authors~\cite{FK01a}, was an 
indispensible tool in our quantitative analysis. 

Less accurately known in Eq.~(\ref{nu-flux}) are the first two ingredients, 
the flux of UHEC$\nu$s, $F_{\nu_i}(E_{\nu_i},r)$, and the neutrino number 
density $n_{\nu_i}(r)$. 

The former was assumed to have the form 
$F_{\nu_i}(E_{\nu_i},r)=F_{\nu_i}(E_{\nu_i},0)\,(1+z)^\alpha$, where $z$
is the redshift and where $\alpha$ characterizes the source evolution  
(see also~\cite{Y97,Y99}). The flux at Earth, $F_{\nu_i}(E_{\nu_i},0)$,
has been determined by the fit to the UH\-E\-CR data.  
In our analysis we went up to distances $R_0$ (cf.~(\ref{nu-flux})) 
corresponding to redshift $z = 2$ (cf.~\cite{W95}), 
and uncertainties of the expansion rate~\cite{Groom00}
were included. 
 
The neutrino number density $n_{\nu_i}$ has been treated in the following way.
For distances below 100 Mpc we varied the shape of the $n_{\nu_i}(r)$ 
distribution between the homogeneous case and that of $m_{{\rm tot}}(r)$, 
the total mass distribution obtained from peculiar velocity 
measurements~\cite{Dekel99}. In this way we took into account that the 
density distribution of R$\nu$s as hot dark matter (DM) follows the total mass 
distribution; however, with  less clustering. It should be noted that for 
distances below 100 Mpc the peculiar velocity measurements~\cite{Dekel99} 
suggest relative overdensities of at most a factor 2 $\div$ 3, depending on 
the grid spacing. We did not 
follow the unnatural assumption of having a relative overdensity of 
$10^2\div 10^4$ in our neighbourhood, as it was assumed in earlier 
investigations of the Z-burst hypothesis~\cite{FMS99,W99,W98,Y99,BVZ00}. 
Our quantitative results turned out to be rather insensitive to the 
variations of the overdensities within the considered range, whose effect is 
included in our final error bars. For scales larger than 100 Mpc the R$\nu$ 
density was taken according to the big-bang cosmology prediction, 
$n_{\nu_i}=56\cdot (1+z)^3$ cm$^{-3}$.

\begin{figure}[t]
\begin{center}
\epsfig{file=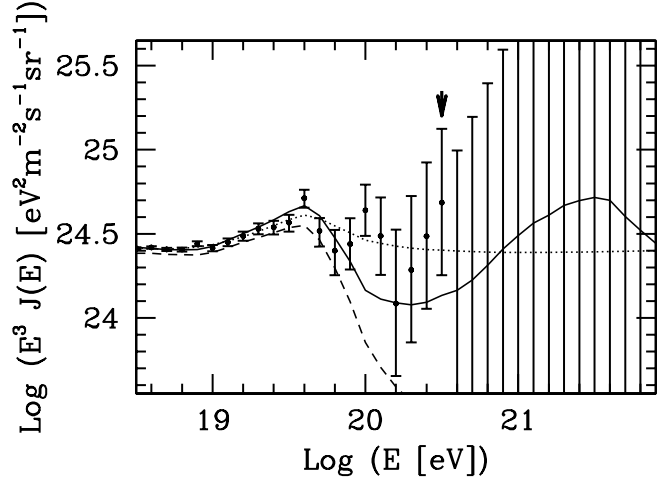,bbllx=20pt,bblly=225pt,bburx=570pt,bbury=608pt,%
width=8.65cm}
\caption[...]{\label{fit_nu}
The available UHECR data with their error bars
and the best fits from Z-bursts~\cite{FKR01a}.
Note that there are no events above $3 \times 10^{20}$~eV
(shown by an arrow). Bins with no events show the 1\,$\sigma$
upper bounds on the flux. Therefore the experimental value of the
integrated flux is in the ``hatched'' region with 68\% confidence level
(``hatching'' is a set of individual
error bars; though most of them are too large to be depicted in full).
The dotted line shows the best fit for the ``halo''-case. The bump around
$4\cdot 10^{19}$~eV is due to the Z-burst protons, whereas the almost
horizontal contribution is the first, power-law-like 
term of Eq.~(\ref{flux}). The solid line shows the ``extragalactic''-case.
The first bump at $4\cdot 10^{19}$~eV represents protons produced at
high energies and accumulated just above the GZK cutoff due to their energy
losses. The bump at $3\cdot 10^{21}$~eV is a remnant of the Z-burst
energy. The dashed line shows the contribution of the power-law-like
spectrum with the GZK effect included. The predicted fall-off for this
term around $4\cdot 10^{19}$~eV can be observed. 
}
\end{center}
\end{figure}

\section{Determination of $m_\nu$ and UHEC$\nu$ flux}

We compared the predicted spectrum~(\ref{nu-flux}) of protons from Z-bursts 
with the observed UHECR spectrum (cf. Fig.~\ref{fit_nu}). 
Our analysis included 
UHECR data
of AGASA~\cite{Takeda98,Takeda99}, Fly's Eye~\cite{Bird93,Bird94,Bird95}, 
Haverah Park~\cite{Lawrence91,Ave00}, 
and HIRES~\cite{Kieda00}. Due to normalization difficulties
we did not use the Yakutsk~\cite{Efimov91} results. 

The predicted number of UHECR events in a bin was taken as
\begin{equation}\label{flux}
N(i)=\int_{E_i}^{E_{i+1}}
{\rm d}E
\left[A \cdot E^{-\beta}+F_Z\cdot j(E,m_\nu)\right],
\end{equation}
where $E_i$ is the lower bound of the $i^{\rm th}$ energy bin. The first
term is the usual power-law behavior, which describes the data 
well for smaller energies~\cite{Takeda98,Takeda99}. For this term we
studied two possibilities. In the first case we assumed
that the power part is produced in our galaxy. Thus no
GZK effect was included for it (``halo''). In the second -- in
some sense more realistic -- case 
we assumed that the protons come from uniformly distributed, EG sources
and suffer from the GZK cutoff (``EG''). In this case the simple
power-law-like term is modified, by taking into account the probability 
$P(r,E_p,E)$, and falls off around $4\cdot 10^{19}$~eV
(see Fig.~\ref{fit_nu}). The second term of the
flux in Eq.~(\ref{flux}) corresponds 
to the spectrum of the Z-bursts, Eq.~(\ref{nu-flux}). 
A and $F_Z$ are normalization factors.
Note that the following implicit assumptions have been made through the
form of Eq.~(\ref{flux}): 
{\em i)} We have assumed (and later checked) that the UHE photons from 
Z-bursts can be neglected. 
{\em ii)} We have assumed that there are no significant additional primary 
UHE proton fluxes beyond the extrapolation of the above power-law. 
This constraint will be relaxed in a future 
publication~\cite{FKR01b}.   

The expectation value for the number of events in a bin is given
by Eq.~(\ref{flux}). 
To determine the most probable value for $m_\nu$ we used the maximum 
likelihood method and minimized~\cite{FK01b} the 
$\chi^2(\beta,A,F_Z,m_\nu)$,
\begin{equation} \label{chi}
\chi^2=\sum_{i=18.5}^{26.0}
2\left[ N(i)-N_{\rm o}(i)+N_{\rm o}(i)
\ln\left( N_{\rm o}(i)/N(i)\right) \right],
\end{equation}
where $N_{\rm o}(i)$ is the total number of observed events in the $i^{\rm th}$
bin. As usual, we divided each logarithmic unit into ten bins. 
Since the Z-burst scenario results
in a quite small flux for lower energies, the ``ankle''  
is used as a lower end for the UHECR spectrum:
$\log (E_{\rm min}/\mbox{eV})=18.5$. Our results are
insensitive to the definition of the upper end (the flux is
extremely small there) for which we choose $\log (E_{\rm max}/\mbox{eV})=26$.
The uncertainties of the
measured energies are about 30\% which is one bin. Using a Monte-Carlo method
we included this uncertainty in the final error estimates.

In our fitting procedure we had four parameters: $\beta,A,F_Z$ and 
$m_\nu$. The minimum of the $\chi^2(\beta,A,F_Z,m_\nu)$ function is 
$\chi^2_{\rm min}$
at $m_{\nu\, {\rm min}}$ which is
the most probable value for the mass, whereas
$\chi^2(\beta',A',F_Z',m_\nu)\equiv \chi^2_o(m_\nu)=\chi^2_{\rm min}+1$
gives the 1\,$\sigma$ (68\%) confidence interval for $m_\nu$.
Here $\beta'$, $A'$, $F_Z'$ are defined in such a way that the 
$\chi^2(\beta,A,F_Z,m_\nu)$ function is minimized in $\beta,A$
and $F_Z$ at fixed $m_\nu$.

Qualitatively, our analysis can be understood in the following way.
In the Z-burst scenario a small R$\nu$ mass needs large 
$E_\nu^{\rm res}$ in order to produce a Z. Large $E_\nu^{\rm res}$ results in 
a large Lorentz boost, thus large $E_p$. In this way the {\em shape} of the
detected energy ($E$) spectrum determines the mass of the R$\nu$. 
The sum of the necessary UHEC$\nu$ fluxes was then determined from the 
obtained {\em normalization} $F_Z$.

Our best fits to the observed data can be seen in 
Fig.~\ref{fit_nu}, for evolution parameter $\alpha=1$. We found a neutrino 
mass of
$2.34^{+1.29(3.74)}_{-0.84(1.66)}$~eV for the ``halo''-  
and $0.26^{+0.20(0.50)}_{-0.14(0.22)}$~eV
for the ``EG''-case, respectively. The first numbers are
the 1\,$\sigma$, the numbers in the brackets are the
2\,$\sigma$ errors. This gives an absolute lower bound on the mass of the 
heaviest $\nu$ of $0.06$ eV at the 95\% CL. The fits are
rather good; for 21 non-vanishing bins and 4 fitted parameters
they can be as low as $\chi^2=15.1$. We determined $m_\nu$ for a wide range
of cosmological source evolution ($\alpha =0\div 3$) and Hubble 
parameter ($H_0=0.64\div 0.78$~km/sec/Mpc) and  
observed only a moderate dependence on them. The results
remain within the above error bars. 
We performed a Monte-Carlo analysis studying higher statistics. In the near 
future, Auger~\cite{Guerard99,Bertou00} will provide a ten times 
higher statistics, which 
reduces the error bars in the neutrino mass to $\approx$ one third of their 
present values.  

\begin{figure}[t]
\begin{center}
\epsfig{file=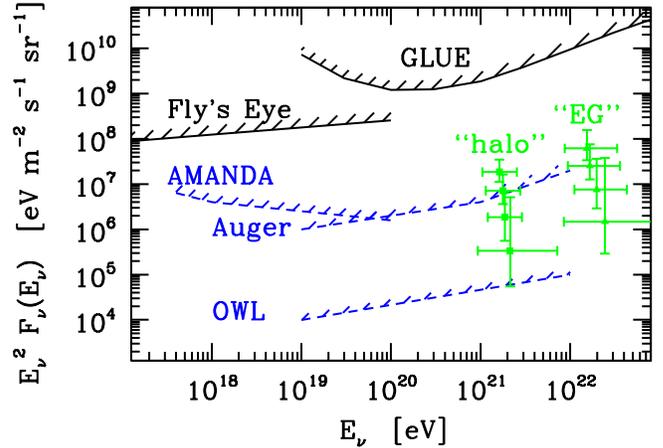,bbllx=20pt,bblly=221pt,bburx=570pt,bbury=608pt,%
width=8.65cm}
\caption[...]{\label{eflux}
Differential neutrino fluxes required by the Z-burst hypothesis for 
the ``halo'' and the ``extragalactic'' case~\cite{FKR01a}. The 
evolution parameter $\alpha$ takes values 0,1,2,3   
from top to bottom for both cases, respectively. 
The horizontal errors indicate the 1\,$\sigma$ uncertainty of the
mass determination and the vertical errors include also the uncertainty
of the Hubble expansion rate.
Also shown are upper limits from Fly's Eye~\cite{Baltrusaitis85} and the 
Gold\-stone lunar ultrahigh energy neutrino ex\-pe\-ri\-ment 
GLUE~\cite{Gorham01}, 
as well as projected sen\-si\-tivi\-ties of AMAN\-DA~\cite{Barwick00}, 
Auger~\cite{Y99,Capelle98} and OWL~\cite{Y99,Ormes97}.}
\end{center}
\end{figure}

The necessary UHEC$\nu$ flux at $E_\nu^{\rm res}$ has been obtained 
from our fit values of the normalization $F_Z$. 
We have summarized them in Fig.~\ref{eflux}, together with some
existing upper limits and projected sensitivities of 
present, near future and future observational projects. 
It is apparent that the flux determination depends much more on the
evolution uncertainties than the mass determination. 
The necessary $\nu$ flux appears to be well below present upper limits
and is within the expected sensitivity of AMANDA, Auger, and OWL.

\section{Comparison with $\triangle m_\nu^2$ from $\nu$ oscillations}

One of the most attractive patterns for $\nu$ masses is similar to
the one of the charged leptons or quarks: the masses are 
hierarchical, thus the mass difference between the families is 
approximately the mass of the heavier particle. Using the 
mass difference of the atmospheric $\nu$ oscillation for the
heaviest mass~\cite{Groom00}, one obtains values between 0.03 and 0.09~eV.  
It is an intriguing feature of our result that the smaller
one of the predicted masses is compatible on the $\approx$ 1.3\,$\sigma$ level 
with this scenario. 

\begin{figure}
\begin{center}
\epsfig{file=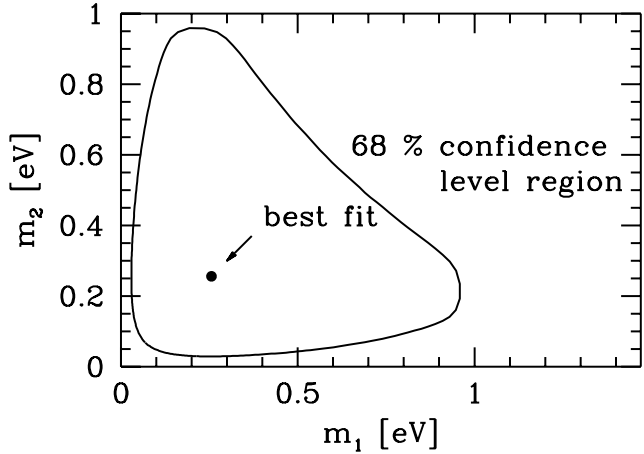,bbllx=20pt,bblly=221pt,bburx=570pt,bbury=608pt,%
width=8.65cm}
\caption[...]{\label{masses}
{
The best fit and the 1\,$\sigma$ (68\% confidence level) region in a scenario
with two non-degenerate $\nu$ masses~\cite{FKR01a}.
}}
\end{center}
\end{figure}

Another popular possibility is to have 4 neutrino types. Two of them 
-- electron and sterile neutrinos -- are
separated by the solar $\nu$ oscillation solution, the other two
-- muon and tau -- by the atmospheric $\nu$ oscillation 
solution, whereas the mass difference between the two groups is
of the order of 1~eV. We studied this possibility, too. On our
mass scales and resolution the electron and sterile neutrinos are 
practically degenerate with mass $m_1$ and the muon and tau
neutrinos are also degenerate with mass $m_2$. The best fit and the  
1\,$\sigma$ region in the $m_1-m_2$ plane  
is shown in Fig.~\ref{masses} for the ``EG''-case. 
The dependence of this result on the cosmological evolution and 
on the UHEC$\nu$ spectrum will be discussed elsewhere~\cite{FKR01b}.
Since
this two-mass scenario has much less constraints the allowed region 
for the masses is larger than in the one-mass scenario.   

\section{Differences with respect to Yoshida {\em et al.} (1998)}

Numerical simulations of Z-burst cascades for $m_\nu\sim 1$ eV, 
taking into account all known EG propagation effects, were performed
by Yoshida {\em et al.} \cite{Y99}. Based on case studies, relative 
overdensities 
of R$\nu$s ranging from $20\div 10^3$ 
on a scale of 5 Mpc were argued to be necessary in order to 
get a successful description of the UHECR events and rate above the
GZK cutoff without violating lower energy photon flux limits and
without invoking inconceivable UHEC$\nu$ fluxes. For such large overdensities,
most of the UHECRs from Z-bursts originate nearby and their 
attenuation to the Earth can be neglected. In our case, with realistic
overdensities $\leq 2\div 3$ on scales $\leq 100$ Mpc, most of 
the UHECRs from Z-bursts originate from cosmological distances. Therefore, 
despite of the fact that by 
construction the overall rate of UHECRs from Z-bursts observed at Earth 
is the same in both investigations, the predicted spectra 
are quite different. No large overdensity is needed to reproduce the
data. Note that the EG scenario is dominated not by the nearby 
Z-burst but by the pile-up of Z-burst protons due to the GZK 
effect (cf. Fig.~\ref{fit_nu}).   

\section{Conclusions}

We reported on a comparison of the predicted spec\-trum of the 
Z-burst hypothesis with the observed
UHE\-CR spectrum~\cite{FKR01a}. 
The mass of the heaviest R$\nu$ turned out to be 
$m_\nu=2.34^{+1.29}_{-0.84}$ eV for halo and
$0.26^{+0.20}_{-0.14}$~eV for EG scenarios. 
The second mass, 
with a lower bound of $0.06$ eV on the 95\% CL, is 
compatible with a 
hierarchical $\nu$ mass scenario with the largest mass suggested by 
the atmospheric $\nu$ oscillation. 
The above $\nu$ masses are in the range 
which can be explored by future laboratory experiments
like the $\beta$ decay endpoint spectrum  and the 
$\nu$ less $\beta\beta$ decay~\cite{PW01,FPS01}.
They compare also favourably with the tau(?) neutrino mass range 
$0.04\div 4.4$ eV found recently from a detailed analysis of the
latest CMBR measurements~\cite{WTZ01}.  
We analysed a possible
two-mass scenario and gave the corresponding confidence
level region. The necessary UHEC$\nu$
flux was found to be consistent with present upper limits
and detectable in the near future.

\begin{acknowledgements}
We thank S. Barwick, 
O. Biebel, S. Bludman, W. Buchm\"uller, K. Mannheim, H. Meyer, 
W. Ochs, and K. Petrovay for useful 
discussions. 
We thank the OPAL collaboration for their unpublished
results on hadronic Z decays.
This work was partially supported by Hungarian Science Foundation
grants No. OTKA-\-T34980/\-T29803/\-T22929/\-M28413/\-OM-MU-708. 
\end{acknowledgements}

\end{document}